# Machine learning interatomic potential for the low-modulus Ti-Nb-Zr alloys in the vicinity of dynamical instability

Boburjon Mukhamedov[1,*], Ferenc Tasnádi[1] and Igor A. Abrikosov[1]

[1] *Theoretical Physics Division, Department of Physics, Chemistry and Biology (IFM), Linköping University, SE-581 83, Linköping, Sweden*

[*] Corresponding author: boburjon.mukhamedov@liu.se

## Abstract

Machine learning-augmented first-principles simulations facilitate the exploration of alloying and thermal treatments for tailoring material properties in industrial applications. However, addressing challenges near dynamical instabilities requires rigorous validation of machine-learned interatomic potentials (MLIP) to ensure their reliable applicability. In this study we have trained MLIP using moment tensor potentials to simulate finite temperature elastic properties of multicomponent β-$Ti_{94-x}Nb_xZr_6$ alloys. Our simulations predict the presence of the elinvar effect for the wide range of temperatures. Importantly, we predict that in a vicinity of dynamical and mechanical instability, the β-$Ti_{94-x}Nb_xZr_6$ alloys demonstrate strongly non-linear concentration-dependence of elastic moduli, which leads to low values of moduli comparable to that of human bone. Moreover, these alloys demonstrate a strong anisotropy of directional Young's modulus which can be helpful for microstructure tailoring and design of materials with desired elastic properties.

## 1 Introduction

Simulating systems like metals and alloys with dynamical instability at zero temperature presents several challenges due to the complex, often unpredictable behaviors of these materials under various conditions [1]. Dynamically unstable systems exhibit fluctuating atomic positions, phase transitions, and other non-trivial effects that are difficult to capture accurately either in theoretical simulations or in physical experiments. Classical interatomic potentials often lack the flexibility to model interatomic interactions in these systems, as they assume relatively stable bonding patterns. Accurate predictions of properties of such systems requires carrying out the ab-initio molecular dynamics simulations (AIMD). Traditional quantum mechanical methods like Density Functional Theory (DFT) provide accurate descriptions of atomic interactions but are computationally expensive, especially for large systems or long simulation times. Simulating the wide range of alloy compositions and temperatures will further increase the computational cost. To overcome this issue, one needs to consider alternative methods.

In recent years the machine learning interatomic potentials (MLIPs) have gained a lot of popularity as a powerful tool that brings together the high accuracy of quantum mechanical methods and the efficiency of classical interatomic simulations [2-12]. A key component of MLIPs is how atomic environments are described. The atomic positions and local environments must be converted into a form the machine learning model can understand. Among the widely used MLIPs, Neural Network Potentials, like the Behler-Parrinello potential [4,13-15], pioneered MLIPs by capturing complex atomic interactions using symmetry functions. The Gaussian Approximation Potential [16,17], which leverages Gaussian process regression and Smooth Overlap of Atomic Positions descriptors, is another widely used model known for its flexibility across diverse materials, including alloys and molecular systems [18-20]. Physically Informed Neural Networks [21-23] represent an emerging class of MLIPs that incorporate physical principles directly into the learning process, increasing the model's transferability across conditions not explicitly seen in training. Graph Neural Networks (GNNs) [24,25] offer another cutting-edge approach by directly modeling atomic environments without predefined descriptors. These GNN-based MLIPs are particularly effective in complex molecular systems and dynamic simulations, pushing the boundaries of accurate, large-scale atomistic modeling across materials science and chemistry [26]. Atomic Cluster Expansion (ACE) [10] provides a systematic and physics-informed representation of potential energy surface using polynomial-like cluster functions. MACE [27], a machine-learning-augmented ACE, incorporates neural networks to improve accuracy while maintaining physical constraints like rotational and translational symmetry.

Moment Tensor Potentials (MTPs) [28,29] have become prominent for their efficient modeling of complex atomic environments, particularly in metals, alloys, and ceramics. MTPs employ a unique tensor-based descriptor that systematically captures interatomic forces and energies with high precision, making them computationally efficient for large systems. They excel in simulating materials under varying conditions, such as high pressure or temperature, and are particularly well-suited to model systems with diverse atomic species. It was proven to effectively predict the thermodynamic and physical properties of materials with the same accuracy as AIMD [30,31]. Shapeev et al [30] showed that for bcc Ti (the high-temperature phase), the MTP predicted temperature variation of elastic constants are in good agreement with that of AIMD. For (Ti-Al)N alloys, it was also reported that MTP predictions of elastic properties are in very good agreement with AIMD result [31]. However, to the best of our knowledge, the excellent performance of MTPs was demonstrated in dynamically stable systems.

In this work, on the example of β-type Ti alloys we extend the use of MTP-MLIP towards predicting the materials properties in a vicinity of dynamical instability. Pure Ti has two allotropic phases: α- and β-phase. The α-Ti has hexagonal close-packed (hcp) structure and is stable below 1155 K (882 °C), while the β-Ti with body centered cubic (bcc) structure is dynamically unstable at 0 K, as indicated by imaginary frequencies in its phonon dispersion relations [1]. However, the bcc phase of Ti becomes dynamically stable at elevated temperatures, and thermodynamically stable above 1155 K (882 °C). Titanium alloys come in α, β and (α+β) microstructures. Addition of elements such as Mo, Nb, Ta, V and Zr can stabilize β-phase [32-38], while addition of Al, N and O stabilize α-phase. The dual phase (α+β) alloys such as Ti-6Al-4V-based solid solutions are widely used for aerospace applications [39,40], e.g. fan blades in jet engine. The β-type Ti alloys are more ductile and largely utilized in the biomedicine field, since for most of these applications the elastic modulus of alloys needs to closely match to that of human bone to avoid the bone degradation. They are used as materials for artificial joints, screws, plates for fractures, and for other biological implants. The main problem for the design of implants is to find a material that is compatible with the bone tissue and has elastic moduli 30 GPa–50 GPa. Currently for these applications the β-Ti alloys with addition of Nb, Ta, Zr, Mo, Sn and Hf are utilized [41-44]. However, it is important to note that β-Ti alloys remain metastable in terms of thermodynamics [45,46], meaning that despite the presence of alloying elements, they do not represent the lowest energy state and may undergo phase transformations under certain conditions, such as deformation or aging. These transformations include spinodal decomposition or formation of α and ω phases [47,48].

Despite the technological importance of Ti alloys, their theoretical understanding is still limited, e.g. because of challenges of simulating Ti-rich bcc alloys due to their dynamical instability at low temperature [49]. Considering elastic properties of Ti-based alloys, Raabe et al [50] carried out first-principles simulations of stability and elastic moduli of binary Ti-Nb alloys. According to this study, increased Nb content helps to stabilize bcc structure but also increases the elastic modulus of alloys. Similar results were reported by Karre et al [51] for Ti-Nb and Ti–Nb–Zr alloys. Hu et al [52] and Dai et al [53] studied how alloying elements can influence the phase stability and Young's modulus in the Ti alloys. Skripnyak et al [54] calculated the elastic properties of β-type Ti-V alloys, focusing on the alloy composition in the vicinity of the mechanical instability. These calculations were verified by experimental measurements of Young's modulus. In Ti-rich regions, the Ti-V alloys were mechanically unstable, $C' < 0$, but increasing V content stabilized bcc structure [54]. High anisotropy of Young's modulus was also predicted in Ti-V alloys close to the point of instability [54]. Calculations of elastic properties for bcc Ti-V alloys across the full concentration range were performed in [55]; however, this study was limited to $T$ = 0 K. Huang et al [56] reported a first-principles study of temperature effects on relative phase stability of α, β, and ω phases of Ti alloys with addition of 3d, 4d, and 5d group elements. However, there is a very few theoretical studies on the influence on effect of finite temperatures on the elastic properties of Ti alloys.

In this work, we train the MTP and apply it to accurately describe the elastic properties of β-type $Ti_{94-x}Nb_xZr_6$. We investigate the solid solutions with different concentrations of Nb and for the temperature interval from 300 up to 1300 K. As noted above, for some Ti-rich compositions, bcc solid solutions may not be the thermodynamically stable phases [45-47]. However, the single-phase metastable bcc alloys can be synthesized for similar systems [41-44]. This motivates us to identify the regions of mechanical and dynamical instability in the β-type $Ti_{94-x}Nb_xZr_6$ alloys. Close to these regions, we observe a sensitive behavior of the elastic properties with variations of temperature and composition. We predict high anisotropy of Young's modulus close to the point of the instability which provides additional opportunity to tune the elastic properties of the alloys, useful for the design of β-Ti alloys with desirable mechanical properties.

## 2 Methodology

### 2.1 DFT calculations

Theoretical calculations of $Ti_{94-x}Nb_xZr_6$ alloys were carried out in the framework of DFT using projector augmented wave (PAW) [57] method implemented in Vienna ab-initio software package (VASP) [58,59]. Chemical disordered of $Ti_{94-x}Nb_xZr_6$ alloys was simulated using special quasi-random structures (SQS) method [60,61]. The SQS supercells provide adequate structural models for alloys synthesized by rapid quenching from high temperatures. Five different Nb concentrations have been considered: 1, 6, 15, 22 and 30 at.%. For each composition, we generated 4x4x4 unit cell sized SQS supercells containing 128 atoms. Generalized gradient approximation (GGA) [62] was used to describe the exchange and correlations effects. The cutoff energy for plane waves was set to 700 eV. The relaxation of atomic positions was carried out by calculating the Hellman-Feynman forces [63] and the stress tensor and using them in the conjugated gradient method. The convergence criterion for electronic subsystem was set to $10^{-4}$ eV/atom for subsequent iterations. Integration in the reciprocal space was performed over a grid of 4x4x4 k-points.

## 2.2 Moment tensor potential

The MTP method was employed to fit a potential capable of predicting the thermodynamic and mechanical properties of $Ti_{94-x}Nb_xZr_6$ alloys. This method, based on the force field approach, uses linear regression to map atomic configurations to a set of basis functions. For this work, we selected the 16g MTP with 380 basis functions, and set the cutoff distance to 5 Å, which is the recommended value that balances the trade-off between computational cost and accuracy [29]. The choice of the 16g MTP was guided by prior studies [31, 64], where it was demonstrated that this potential provides accurate and reliable predictions of the $C_{ij}$ elastic constants for $Ti_{0.5}Al_{0.5}N$. Verification of accuracy of the MTP potentials obtained with the above settings is presented in Supplementary Information, Sec. S1 and S2.

## 2.3 Initialization of MTP

In this work we start with a workflow developed by F. Bock et al [31] for the training of MTP for dynamically stable alloys. The initial MTP was trained on a low-accuracy dataset generated from Γ-point-only AIMD simulations. This significantly reduces the computational resources required to generate initial training set (TS). The AIMD were performed using the NVT canonical ensemble for $Ti_{94-x}Nb_xZr_6$ alloys with 15, 22 and 30 at.% Nb and four different temperatures: 700, 900, 1100 and 1300 K. Within these temperature and composition ranges, the alloys were expected to be dynamically stable. This was done to prevent unphysical configurations in the initial TS. The approximate volume of each $Ti_{94-x}Nb_xZr_6$ supercell was determined using experimental atomic volumes of pure Ti, Nb and Zr, and by estimating the alloys volume based on its atomic composition. AIMD simulations were run for up to 10000 timesteps, and to ensure equilibrium, the first 3,000 timesteps were excluded. For each composition, the next 5,000 timesteps from all temperatures were combined into a training set, while the remaining data formed the validation set. This resulted in three training sets and three validation sets corresponding to $Ti_{94-x}Nb_xZr_6$ alloys with 15, 22, and 30 at.% of Nb.

The MTP initialization involved an active selection procedure with multiple iterations of training to improve the interatomic potential. For the first MTP, we randomly selected 150 atomic configurations from the three TSs (50 configurations from each TS). Then, the extrapolation grades $\gamma$ of configurations in the TSs were evaluated. From configurations with $\gamma$ values exceeding a specific selection threshold, we selected 30 new datapoints (10 configurations from each TS) and added them to the first 150 datapoints, followed by refitting the MTP with this updated dataset. This process constituted the first iteration of active selection. As iterations progressed, the selection threshold was reduced. Initially set to $\gamma = 150$, iterations continued until the extrapolation grades in the TSs were $\gamma < 3.0$, which took 10 iterations. At the end of the 10th iteration, the dataset contained 420 configurations.

## 2.4 Active learning of MTP

The MLIP-2 code allows the integration of a trained MTP with LAMMPS software [65], enabling MD simulations where the trained MTP predicts the energy, forces, and stresses for each configuration at each MD time step. It is important to note that the initial MTP was trained on a low-quality dataset, which requires replacement with higher-accuracy data to improve model fidelity. To enhance dataset and potential quality, we employed an active learning (AL) approach. At each MD time step, the MTP calculates an extrapolation grade (γ), and configurations with γ values above the selection threshold ($\gamma^{select}$) are selected for refitting of the potential.

The LAMMPS-MD simulations were run for 30,000 time steps. To prevent the selection of non-physical configurations, we set an additional breaking threshold ($\gamma^{breaking}$ ). For configurations with $\gamma^{select} < \gamma < \gamma^{breaking}$, we performed static DFT calculations with higher accuracy, by using a 4x4x4 k-point grid. These new DFT results were added to the existing dataset, and a new MTP was subsequently trained. This process constitutes a single iteration of AL.

MD simulations were conducted at four temperatures, 700, 900, 1100, and 1300 K, and at zero pressure using the NpT isothermal-isobaric ensemble. In the AL process, we considered a broader range of Nb concentrations in $Ti_{94-x}Nb_xZr_6$: 1, 6, 15, 22, and 30 at.% Nb. During 30 AL iterations, the potential's predictive accuracy gradually improved. It should be noted that the training dataset initially included both low-quality data (from Γ-point-only AIMD) and higher-quality data from AL. To further refine the MTP, all low-quality data were removed, followed by an additional 25 iterations of AL to replace these configurations.

AL iterations were concluded once the extrapolation grades for all new configurations remained below $\gamma < 1.2$ over 30,000 MD time steps. To evaluate the MTP's performance at lower temperatures, we conducted additional NpT-MD simulations at 300 and 500 K and calculated the γ values for the new configurations. For the low-Nb $Ti_{93}Nb_1Zr_6$ alloy at 300 K, there were aroud 200 configurations (out of 30,000 MD time steps) with γ values of 1.2 – 2.8, which are exceeding those in higher-temperature MD simulations. For other Nb concentrations, γ remained below 1.2 at lower temperature MD simulations. Therefore, an additional five AL iterations were performed for the $Ti_{93}Nb_1Zr_6$ alloy at low temperatures, incorporating new high-γ configurations into the training set. After these iterations, the final MTP maintained $\gamma < 1.2$ over 30,000 MD time steps for all Nb concentrations and the temperature range of 300 to 1300 K. The final dataset comprised 880 configurations.

The requirement for additional AL iterations for $Ti_{93}Nb_1Zr_6$ can be expalined by its dynamical instability at low temperatures, which we will disscuss in more details in Section 3.2. It is important to emphisize that the MD runs at 300 and 500 K for low-Nb alloys did not crash or result in unphysical configurations. This "stability" is likely due to the initial training of MTP, followed by AL, was conducted for the compositions and the temperatures where the alloys were expected to be stable. Figure S4 in Supplementary Information shows the average structures of $Ti_{94-x}Nb_xZr_6$ SQSs derived from MD simulations at 300 and 1300 K. One can notice that average structures for low-Nb alloys at 300 K exhibit quite large atomic displacements from ideal bcc positions, whereas higher Nb alloys show minimal displacements. Interestigly, at 1300 K, the avarage structures of all five compositions demonstrate atomic positions very similar to those in ideal bcc lattice. This observeation may also explain the requirement for additional AL iterations for the low-Nb alloys at low temperature.

After completing the active learning for the MTP, we used it to conduct NpT-MD simulations to evaluate the thermal expansion of the alloys within a temperature range of 300 to 1300 K and at zero pressure. Following this, we calculated the elastic constants $C_{ij}$ via stress-strain relationships using the NVT simulations. The strain values were set to ±0.02 and ±0.04. For both NpT and NVT simulations, LAMMPS-MD was run for 30,000 timesteps, with the initial 5,000 timesteps excluded from property estimations to ensure the systems had reached equilibrium.

We then calculated the polycrystalline moduli of the alloys using the Voigt [66], Reuss [67], and Hill [68] averaging methods (see Supplementary Information, Sec. S7 and S8 for details). The directional Young's modulus was determined according to the method of Wortman and Evans [69,70]. To assess the mechanical stability of $Ti_{94-x}Nb_xZr_6$ alloys, we used Born's criteria [71].

$$C_{11} + 2C_{12} > 0;\ C_{11} - C_{12} > 0 \text{ and } C_{44} > 0 \qquad (7)$$

## 3 Results

### 3.1 Elastic constants at zero temperature

Before discussing the elastic properties of Ti-Nb-Zr alloys at finite temperatures, we first verified whether the trained MTP can reproduce the PAW calculations at $T$ = 0 K. Figure 1. compares the elastic properties at $T$ = 0 K obtained from PAW calculations with the predictions using MTP. To calculate the elastic constants $C_{ij}$, we used the stress-strain relation, and the strain values were set to ±0.02 and ±0.04. To properly account for symmetry of alloys in the finite-size SQS calculations, the $C_{ij}$ values were averaged over three orientations which was done by rotating the strain matrix twice: first with a 90° rotation around the Z-axis, and then 90° rotation around the Y-axis [64,72]. This resulted in three sets of $C_{ij}$ values for each supercell, with the reported $C_{ij}$ being the average across these sets. For more details on the $C_{ij}$ avereging see the Supplementary information, Sec. S7. The error bars in Figure 1 reflect the variance from this averaging. The MLIP predictions were conducted using the same supercells and strain values as in the PAW calculations, with error bars similarly representing the averaging variance.

As shown in Figure 1, the MTP adequately reproduces the PAW results for $C_{ij}$ constants. The MLIP slightly underestimates $C_{11}$ by ~2 GPa and slightly overestimates $C_{12}$ by ~3 GPa, but these discrepancies remain within the margin of error. This comparison demonstrates the reliability of the MTP predictions for elastic properties. Both $C_{11}$ and $C_{12}$ constants increase with Nb content, although the effect on $C_{12}$ is less pronounced. The $C_{44}$ constant, however, demonstrates a sign of mechanical instability when it sharply changes in low Nb-content alloys: the large error bar for the $Ti_{79}Nb_{15}Zr_6$ alloy reflects an uncertainty in determining the $C_{44}$.

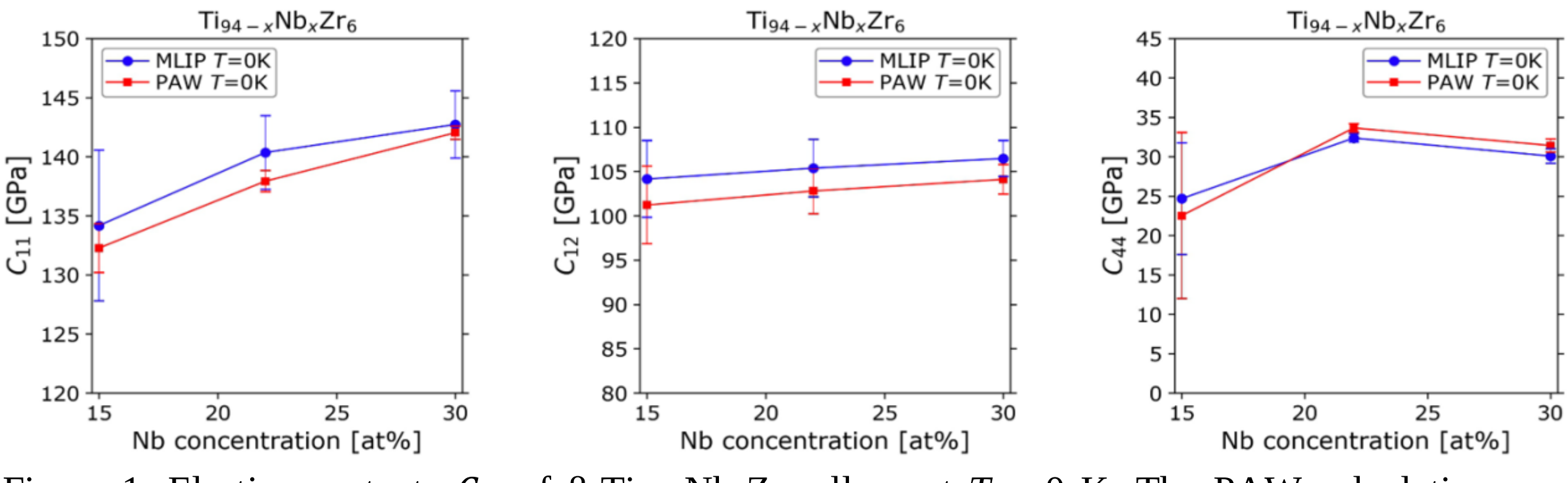

Figure 1. Elastic constants $C_{ij}$ of β-$Ti_{94-x}Nb_xZr_6$ alloys at $T$ = 0 K. The PAW calculations are compared with MLIP predictions. The error bars in elastic constants $C_{ij}$ are from averaging (see text above).

In Figure 1, we did not include the $C_{ij}$ results for the two low-Nb alloys, $Ti_{93}Nb_1Zr_6$ and $Ti_{88}Nb_6Zr_6$, due to their dynamical instability during ionic relaxation at 0 K. Specifically, these low-Nb structures failed to stabilize into a bcc lattice, which is characteristic of β-type Ti-Nb-Zr alloys. As shown in Figure S5 (in Supplementary Information), after relaxation at 0 K and zero pressure, the $Ti_{93}Nb_1Zr_6$ and $Ti_{88}Nb_6Zr_6$ alloys exhibited large atomic displacements, indicating a loss of bcc symmetry. In contrast, high-Nb alloys ($Ti_{79}Nb_{15}Zr_6$, $Ti_{72}Nb_{22}Zr_6$, and $Ti_{64}Nb_{30}Zr_6$) mostly maintained bcc symmetry after relaxation, although the atomic displacements in $Ti_{79}Nb_{15}Zr_6$ are slightly increased (see Figure S5 (c) and (f)). However, to enable a meaningful DFT vs. MLIP comparison across all studied Nb concentrations at 0 K, including low-Nb alloys, we performed DFT calculations on unrelaxed SQS configurations. By avoiding ionic relaxation, we preserved the bcc symmetry in low-Nb alloys. In this case, we minimized total energies and determined the equilibrium volumes for the unrelaxed SQS before calculating the elastic constants.
Figures S6 and Table S1 in Supplementary Information compare DFT and MLIP predictions for $C_{ij}$ elastic constants, lattice parameters, and total energies in unrelaxed SQS configurations. MLIP overestimates $C_{12}$ and underesteemates $C_{44}$. The discrepancies between DFT and MLIP values are approximately: $\Delta C_{11} \approx 2$ GPa, $\Delta C_{12} \approx 5-6$ GPa and $\Delta C_{44} \approx 3-4$ GPa. MLIP slightly underestimates lattice parameters at 0 K, though the difference from DFT predictions is less than 0.1%. The total energy differences between DFT and MLIP predictions at 0 K are approximately 2-4 meV/atom. Overall, we see a good agreement between DFT and MLIP in predicting $C_{ij}$ constants and other properties at 0 K.

### 3.2 Dynamical stability of Ti-Nb-Zr alloys

The properties of Ti-Nb-Zr alloys were investigated across a temperature range from 300 to 1300 K. The thermal expansion of alloys was analyzed using NpT molecular dynamics at zero pressure. Figure 2 shows the temperature dependence of lattice parameters and linear thermal expansion coefficients of β-$Ti_{94-x}Nb_xZr_6$ alloys. The simulation cells for each composition and temperature contain 128,000 atoms. These large cells were created by multiplying the original SQSs, which contained 128 atoms, 10-fold along each of the three crystallographic directions.
In alloys with 1 and 6 at.% Nb, anomalous behavior in the lattice parameters was observed at low temperatures, as indicated by the dashed lines in Figure 2a. This anomality can be attributed to the dynamical instability of these alloys at low temperatures, which will be discussed further below. The coefficient of thermal expansion $\alpha_L$ also increases with Nb content. The $\alpha_L$ results for low-Nb alloys with 1 and 6 at.% Nb are not shown due to the anomalous behavior of their lattice parameters. For comparison, in the gum-metal Ti-23Nb-0.7Ta-2Zr-1.2O, the coefficient of thermal expansion is $0.8 \cdot 10^{-5}$ K$^{-1}$ in the temperature range of 100-700 K [73]. In our calculations for the $Ti_{72}Nb_{22}Zr_6$ alloys, the $\alpha_L$ ~ $0.9 \cdot 10^{-5}$ K$^{-1}$, which is in good agreement with result for the gum-metal.

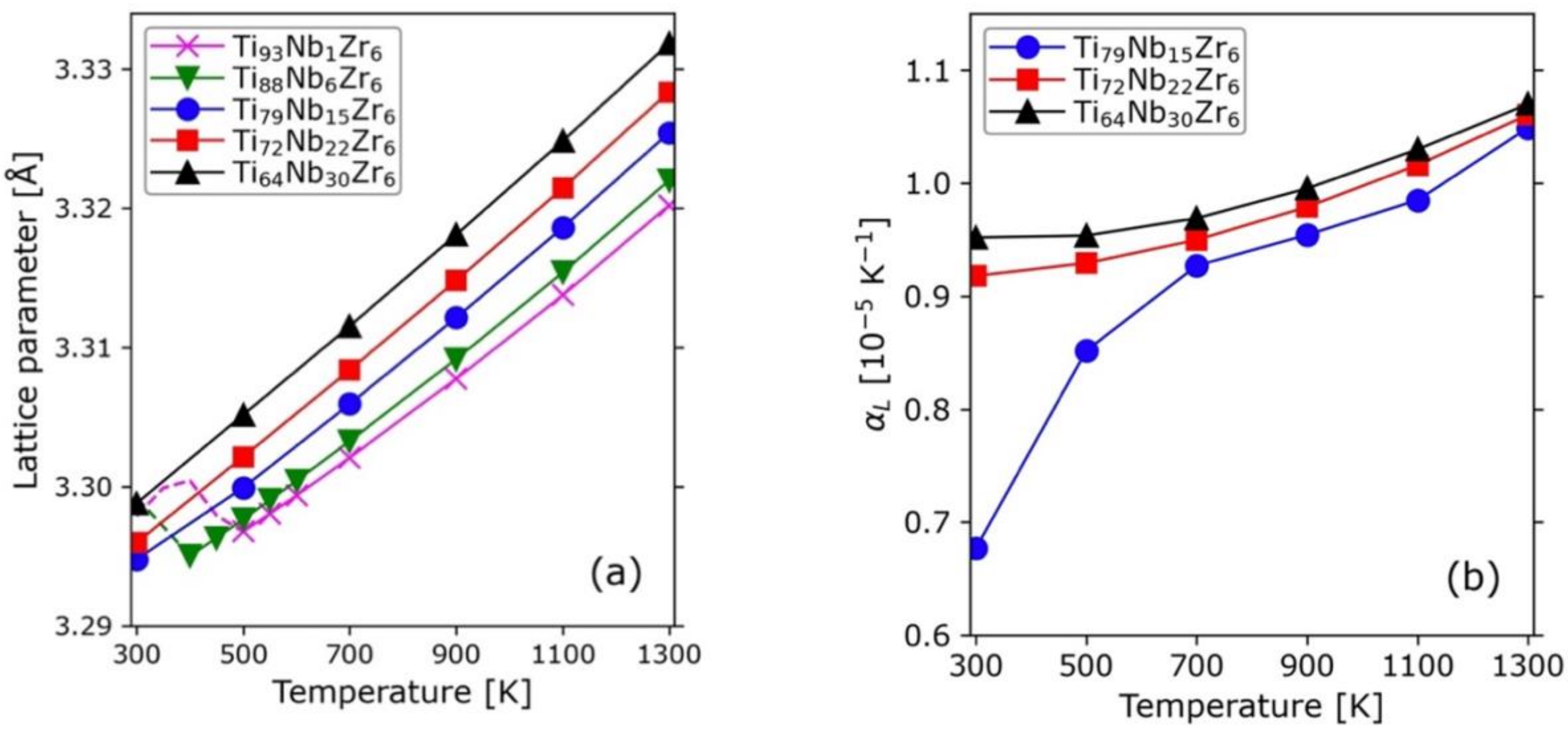

Figure 2. Temperature dependence of (a) lattice parameters and (b) coefficient of linear thermal expansion of $Ti_{94-x}Nb_xZr_6$ alloys with different Nb concentrations. Lattice parameters are obtained from NpT calculations at $P = 0$. Each simulations box contains 128,000 atoms. Dashes lines indicate lattice parameters in regions of dynamical instability.

Next, we used the information on lattice parameters and thermal expansion to further analyze the atomic displacements in $Ti_{94-x}Nb_xZr_6$ alloys. The simulation box for each composition contains 128,000 atoms. Figure 3 shows the mean square displacement (MSD) of atoms from their ideal lattice positions in bcc structure. The MSDs were calculated for each time step of MD simulation, with final value of MSD obtained by averaging over entire MD time steps, excluding the equilibration part of the simulation (first 5000 steps). Thermal expansion of alloys was considered in the evaluation of displacements. The MSD was determined for three crystallographic directions and with respect to lattice parameter which changes with temperature.

As shown in Figure 3, the MSD is expected to increase with temperature. Alloys containing 15, 22 and 30 at.% Nb demonstrate almost linear relationship between MSD and temperature. However, in alloys with lower Nb content (1 and 6 at.%), we observe a deviation from linearity at lower temperatures: after reaching a minimum value, the MSD begins to increase dramatically. The significant magnitude of displacements at low temperatures suggests that the system is attempting to break the bcc symmetry, which is reflected in the tendency of atoms to move further away from their ideal lattice positions. This anomaly in MSD in low-Nb alloys is a sign of dynamical instability, similar to findings of Asker et al [74] for fcc Mo at low temperatures. In that study, it was observed that in fcc Mo the atomic displacements from ideal lattice positions at 300 K, at which it is unstable, were significantly larger than those at 3200 K.

Dinamical instability in the alloys containing 1 and 6 at.% Nb at low temperatures is expected, as the β-phase in pure Ti is also dynamically unstable at low *T* [75,76]. In Figure 3, the temperatures corresponding to the minimum MSD in $Ti_{93}Nb_1Zr_6$ and $Ti_{88}Nb_6Zr_6$ can be considered critical points, below which the alloys become dynamically unstable. We estimate that $Ti_{93}Nb_1Zr_6$ is dynamically unstable below 450-500 K and $Ti_{88}Nb_6Zr_6$ is unstable bellow 350-400 K. Here, we take the highest temperatures as critical point between stable and unstable alloys, therefore, the critical temperatures for $Ti_{93}Nb_1Zr_6$ and $Ti_{88}Nb_6Zr_6$ are ~400 and ~500 K, respectively. These temperature intervals coincide with the observed anomalous behavior of the lattice parameters in the two low-Nb alloys (see Figure 2a).

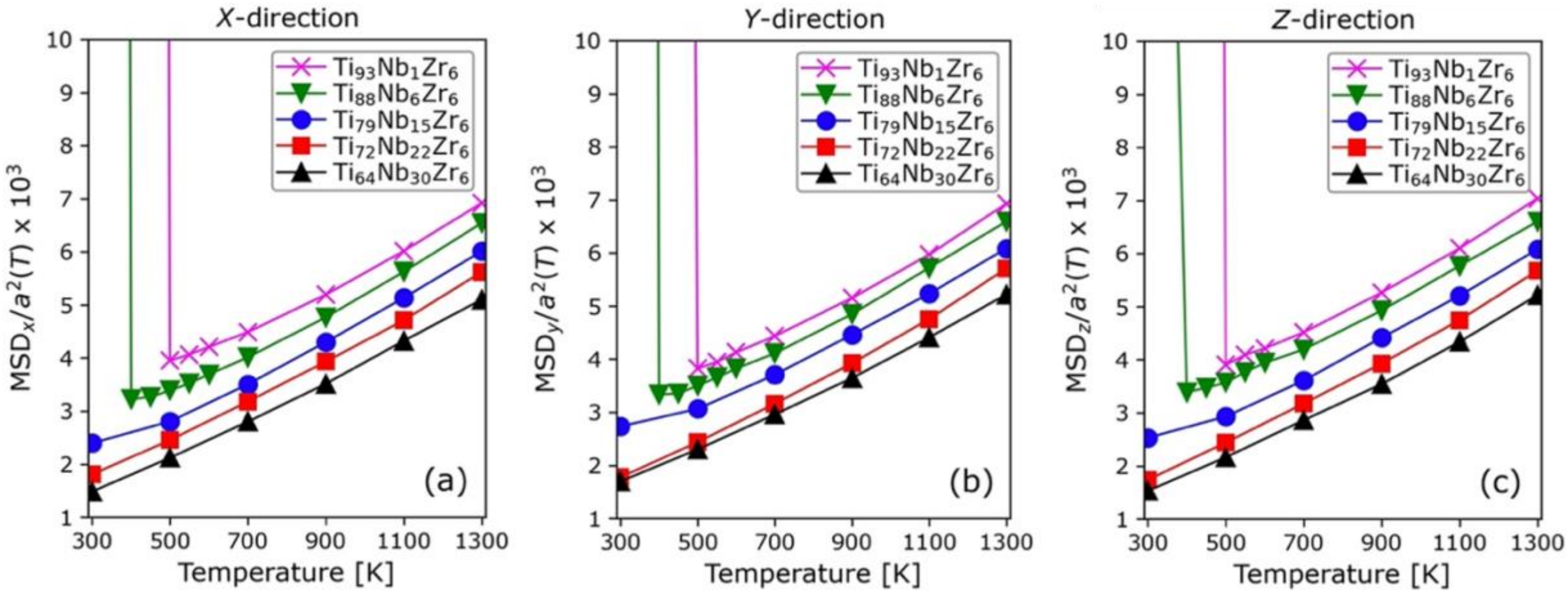

Figure 3. Temperature dependence of mean square displacements MSD of atoms from their ideal lattice positions in $Ti_{94-x}Nb_xZr_6$ alloys. The displacements were determined along three crystallographic directions: (a) $MSD_X$, (b) $MSD_Y$ and (c) $MSD_Z$. The MSDs were determined within NVT simulations, accounting for the thermal expansion of the alloys. Each simulation box contained 128,000 atoms.

Since the MSD shown in Figure 3 correspond to the large supercells containg 128,000 atoms, one might wonder whether the size of the supercell affects the magnitude of displacement or temperatures at which the alloys demonstrate unstable behavior. Figure S7 in Supplementary Information, shows the MSD obtained for supercells containing 128 atoms (original SQSs). Here, we also conduct MD simulations using the trained MTP. Figure S7 indicates that in stable regions the MSD values for smaller cells are not significantly different from those of larger cells. However, the anomalous behavior observed in $Ti_{93}Nb_1Zr_6$ and $Ti_{88}Nb_6Zr_6$ at low $T$ is less pronounced in the smaller supercells. Therefore, to accurately define the instability regions, we used the MSD values from the large supercells.

Another indication of dynamical instability can be observed in a behavior of stress components $S_{ij}$ of undistorted supercells as a function of simulation time. To shows how temperature and Nb content can affect the stresses, in Figure 4 we compare the $S_{ij}$ results for alloys with 1, 6 and 15 at.% Nb at two temperatures, 300 and 500 K. The first 5000 time steps of NVT simulations were excluded from our analysis of the stress components. The thermal expansion of the alloys was taken into account, and for each studied temperature the lattice parameters were obtained through NpT simulations at zero pressure.

According to Figures 4a and 4c, at 300 K the alloys with 1 and 6 at.% Nb exhibit nonzero stresses, indicating instability in within the system. In contrast, for the alloy with a higher Nb content of 15 at.%, the stress components become negligible. Regarding the temperature effect, we observe that at 500 K, the stresses also disappear, which can be clearly seen in case of alloys with 1 and 6 at.% Nb (Figures 4b and 4d). From the MSD results, we determined that for alloy with 1 at.% Nb, the temperature of ~500 K is a border of dynamical instability. The behavior of the stress components strongly correlates with the MSD results shown in Figure 3: increasing the Nb content and temperature can stabilize the $Ti_{94-x}Nb_xZr_6$ alloys, and vice versa, decreasing the temperature and Nb content brings the system closer to dynamical instability.

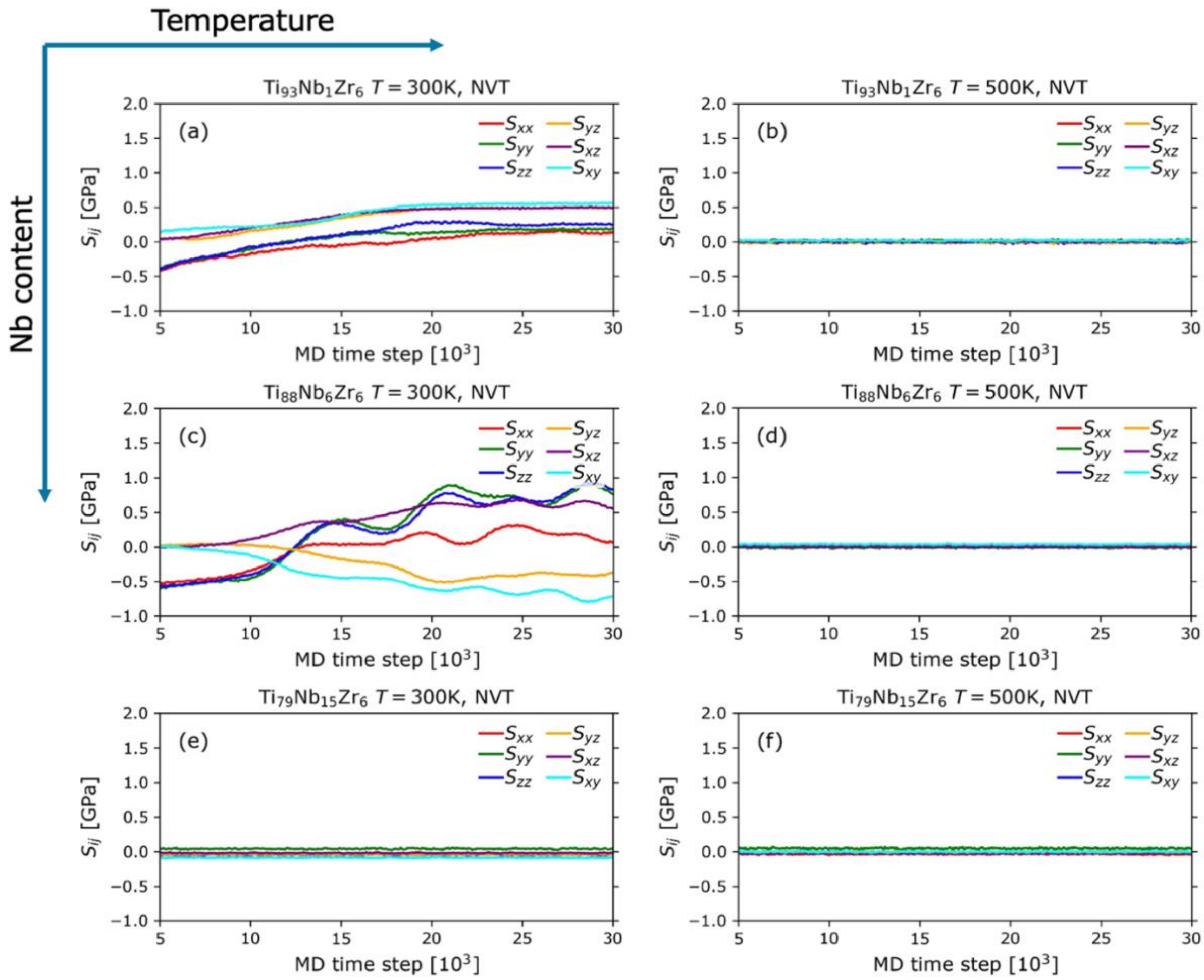

Figure 4. Stress components $S_{ij}$ in undistorted $Ti_{94-x}Nb_xZr_6$ alloys. The $S_{ij}$ are shown for alloys with 1, 6 and 15 at.% Nb and at temperatures of 300 and 500 K. Each simulation box contains 128,000 atoms.

### 3.3 Elastic properties at finite-temperatures

Figure 5 shows the temperature dependence of the calculated $C_{ij}$ of $Ti_{94-x}Nb_xZr_6$ alloys. The error bars are obtained from summarization averaging of constants. The shaded areas in Figures 5a and 5b correspond to the temperature intervals of dynamical instability in $Ti_{93}Nb_1Zr_6$ and $Ti_{88}Nb_6Zr_6$ alloys. Our estimates suggest that $Ti_{93}Nb_1Zr_6$ and $Ti_{88}Nb_6Zr_6$ become dynamically unstable below ~500 and ~400 K, respectively. At low temperatures, both alloys are mechanically unstable, as evidenced by $C_{11} < C_{12}$. Furthermore, within the intervals of dynamical and mechanical instability, a slight softening of the $C_{44}$ constants is also noticeable. In contrast, alloys with higher Nb content do not display any signs of mechanical instability. Within the regions of mechanical/dynamical stability, all five $Ti_{94-x}Nb_xZr_6$ alloys show a weak effect of temperature on the $C_{ij}$ constants: the change in $C_{44}$ constant with temperature is negligibly small for all compositions, and $C_{11}$ and $C_{12}$ constants decrease by 5-10 GPa as they approach the highest studied temperature.

The low-temperature dependence of elastic constants in the studied alloys is not surprising. It is known that bcc Ti-Nb-based alloys can demonstrate an elinvar effect in the wide range of temperatures [73,77]. Saito et al [73] reported that the gum-metal, Ti-23Nb-0.7Ta-2Zr-1.2O alloy, after cold-working demonstrate elinvar and invar behavior across wide temperature range: the elastic modulus and lattice parameter remains about constant between 77 and 500 K. Dubinskiy et al [77] observed elinvar behavior in Ti-22Nb-6Zr (at.%) alloy on cooling from 820 to 420 K. In the work by Saito et al [73] the effects of microstructure evolution were significant for elinvar behavior. On the other hand, Dubinskiy et al [77] reported that the origin of such behavior is not a result of any magnetic or structural phase transformation, change in dislocation density or low-symmetry crystal lattice-related phenomena, but rather of negligible temperature dependences of the elastic constants of β-phase. Our calculations of elastic properties at finite temperature and conclusions made in [77] regarding what causes the elinvar effect in Ti-22Nb-6Zr (at.%) alloy indicate that the peculiar behavior of the alloys is native to bcc phase and is related to its proximity to dynamical instability.

According to Figure 5, both $C_{11}$ and $C_{12}$ constants of $Ti_{94-x}Nb_xZr_6$ alloys slightly decrease with temperature, but the effect of temperature on $C_{12}$ is stronger than in $C_{11}$. Therefore, $C' = (C_{11} - C_{12})/2$ becomes larger with temperature, as shown in Figure 5f. We should note that the MTP simulations for pure β-Ti performed by Shapeev et al [30] also demonstrated that $C_{12}$ has a bit stronger temperature dependence than $C_{11}$, and the $C'$ parameter becomes higher at finite temperatures. In fact, it was also predicted that β-Ti exhibit elinvar effect in the wide range of temperatures from 900 to 1700 K, as seen in a weak temperature dependence of its elastic constants [30].

Figure 5f shows that in $Ti_{94-x}Nb_xZr_6$ alloys the temperature dependence of $C'$ parameter becomes weaker with increased Nb content as the alloys become more mechanically stable. In the alloys with $x_{Nb} \geq 15$ at.% the criteria of mechanical stability $C' > 0$ is fulfilled in the whole range of studied temperatures. As we approach the temperatures and concentrations where $C'$ begins to exhibit a sensitive behavior to rather small deviations, we expect a strong softening of alloys. Notably, low-Nb alloys containing 1 and 6 at% Nb demonstrate a strong temperature-induced strengthening in $C'$, likely because these alloys move away from the instability regions and become "more" mechanically stable at higher temperatures. In fact, at high temperatures, $T > 900$ K, the low-Nb alloys start exhibiting the temperature dependence of $C'$ almost similar to that observed in higher Nb-content alloys, as shown in Figure 5f. Since $C'$ parameter is directly involved in estimation of polycrystalline shear ($G$) and Young's ($E$) moduli, one should expect the $T$-induced strengthening in these moduli as well. In our simulations, the $C'$-strengthening is not a result of a phase transition, but due to alloys (especially with low-Nb) becoming mechanically more stable at higher temperatures.

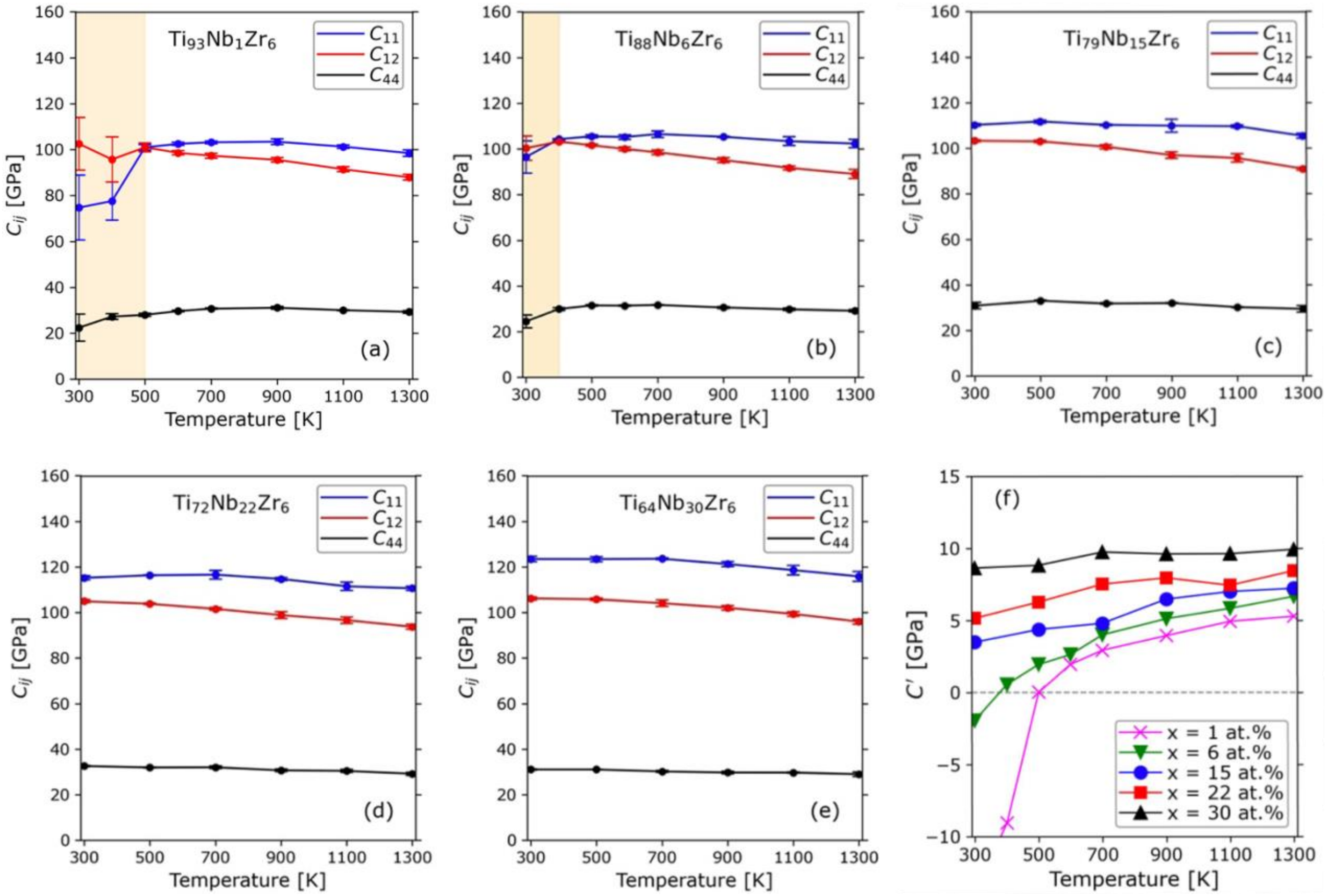

Figure 5. Temperature dependence of (a-e) $C_{ij}$ elastic constants and (f) $C'$ parameter of $Ti_{94-x}Nb_xZr_6$ alloys with different Nb concentrations: 1, 6, 15, 22 and 30 at.%. Shaded areas in (a) and (b) indicate the regions with dynamical instability. The error bars in (a-e) are from averaging of $C_{ij}$.

Let us make few more observations regarding the dynamical and mechanical instability in the low-Nb $Ti_{93}Nb_1Zr_6$ and $Ti_{88}Nb_6Zr_6$ alloys. Both alloys become mechanically unstable within the same temperature intervals as they become dynamically unstable. The calculations of $C'$ constant (see Figure 5f) shows that $Ti_{93}Nb_1Zr_6$ and $Ti_{88}Nb_6Zr_6$ alloys become mechanically unstable below ~500 and ~390 K, respectively, and their dynamical instability also occurs below ~500 and ~390 K, respectively (see Figure 3). Within the scope of this work, we cannot be certain that dynamical and mechanical instabilities coincide. Dynamical instability in materials is often related to negative value of $C'$ in the long-wavelength limit, Γ-point, however in pure β-Ti the dynamical instability also occurs along N-Γ and H-P-Γ branches of wave vectors in the first Brillouin zone [76]. Considering that areas of dynamical and mechanical instabilities in case of $Ti_{93}Nb_1Zr_6$ and $Ti_{88}Nb_6Zr_6$ are very similar, one can assume that they are dynamically unstable at Γ-point.

### 3.4 Polycrystalline moduli at finite-temperatures

Figure 6 shows the temperature dependence of bulk *B*, shear *G* and Young's *E* moduli for $Ti_{94-x}Nb_xZr_6$ alloys calculated using Voight-Reuss-Hill averaging method [66-68]. As expected, alloys with low Nb content exhibit anomalous elastic modulus behavior at low temperatures due to instability. In Figure 6a, the *B*-modulus of alloys decreases by approximately 10 GPa between 300 and 1300 K. Both *G* and *E* moduli follow similar temperature-dependent trends; in low-Nb alloys, they display a nonlinear relationship as the alloys approach instability. Additionally, temperature-induced strengthening is observed in both $G$- and $E$-moduli, similar to that seen in the $C'$ parameter. This behavior was discussed in more detail in the previous section. As shown in Figure 6, increased Nb content improves mechanical stability and increases elastic moduli of alloys. Alloys with 15, 22 and 30 at.% Nb, which are mechanically stable, exhibit a weak temperature dependence of $G$- and $E$-moduli, indicating the presence of elinvar effect across a broad temperatures range. Our simulations suggest that even alloys containing 1 and 6 at.% Nb show a sign of elinvar behavior within the temperature range of their mechanical stability at T > 900 K.

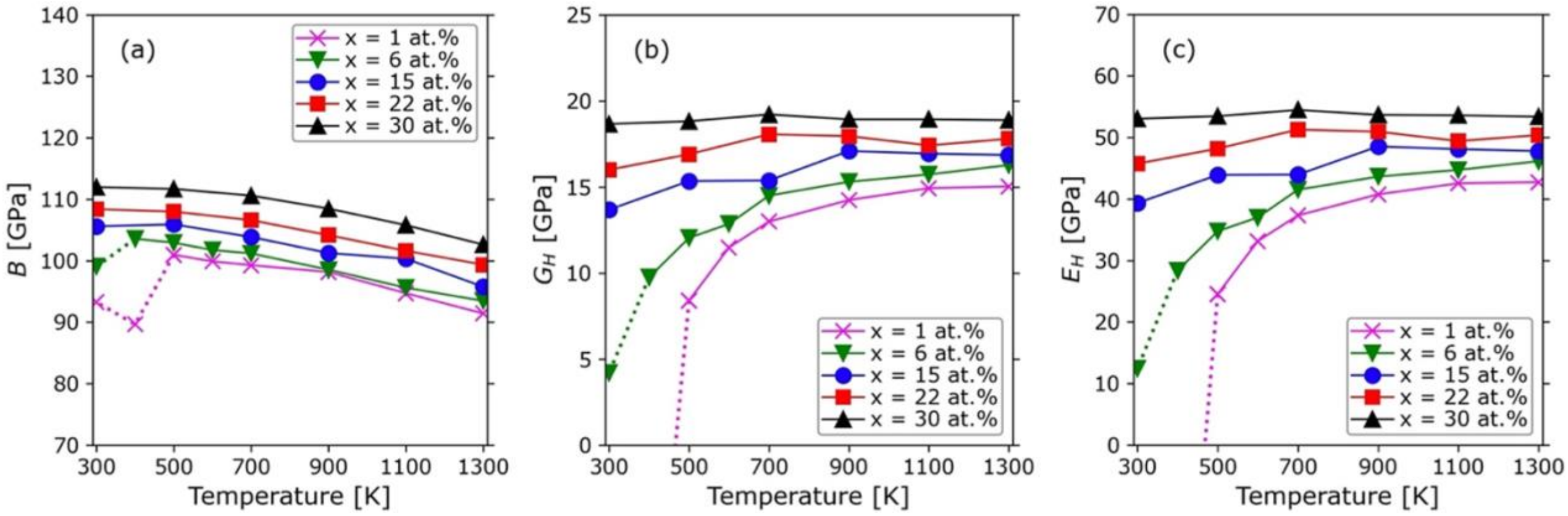

Figure 6. Temperature dependence of polycrystalline moduli *B*, *G* and *E* in five $Ti_{94-x}Nb_xZr_6$ alloys. Dashed lines correspond to the condition of dynamical/mechanical instability when $C' < 0$.

Figure 7 presents the color map of Young's modulus $E_H$ of $Ti_{94-x}Nb_xZr_6$, calculated using the Hill averaging. This map illustrates $E_H$ as a function of temperature and Nb content. The lower left corner indicates a region of dynamical and mechanical instability: the dashed and solid lines mark the borders of dynamical and mechanical instability, respectively. According to Figure 7, at room temperature alloys with $x_{Nb} > 12$ at.% should be both dynamically and mechanically stable. Near the point of instabilities, one can notice a significant softening of $E_H$ modulus. Our assessment shows that β-$Ti_{94-x}Nb_xZr_6$ with Nb content in the range of 12 – 17 at.% exhibit $E_H$-modulus between 30 and 40 GPa. Further increases in Nb content result in a higher elastic modulus.

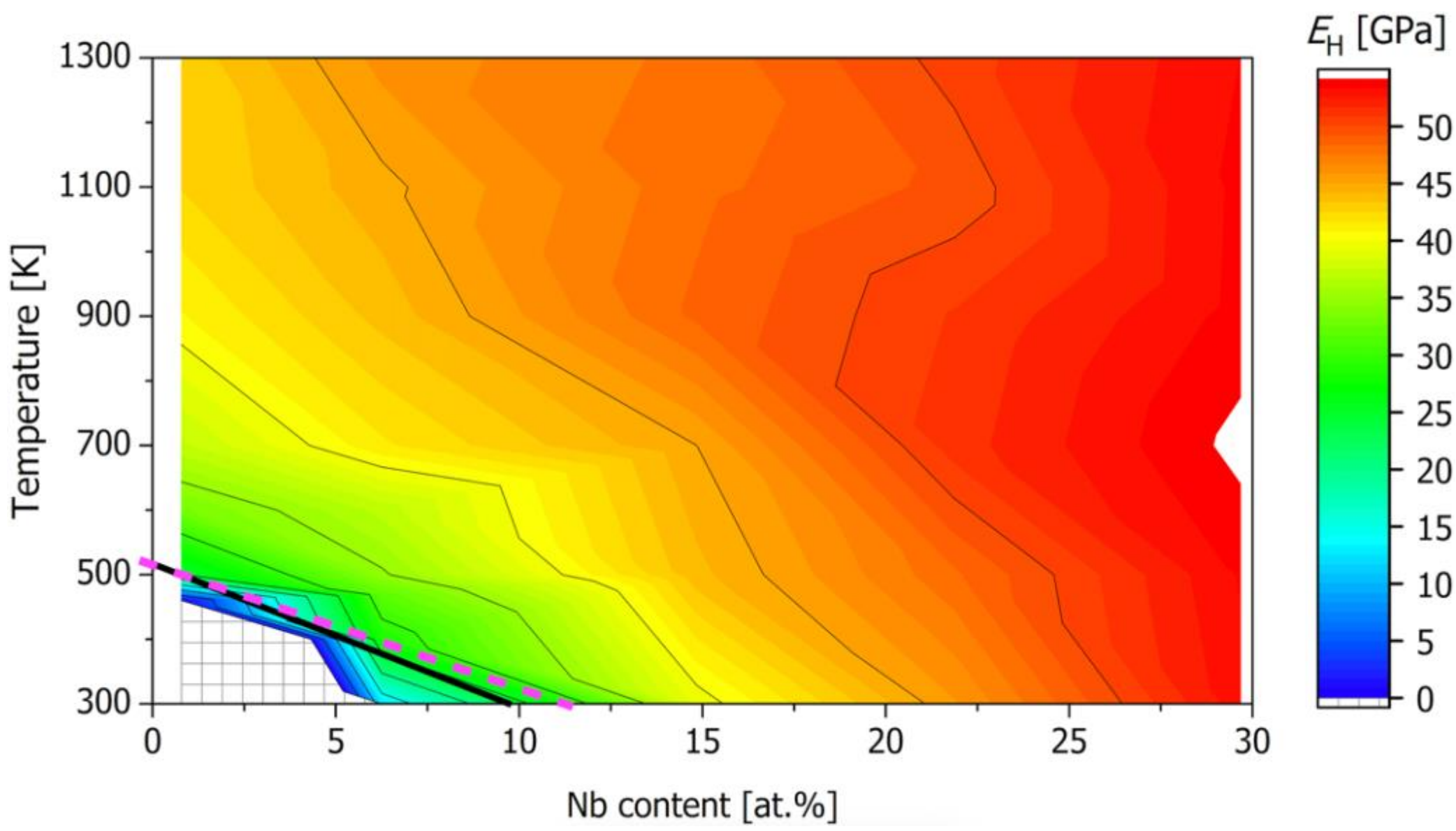

Figure 7. Color map of Young's modulus $E_H$ of $Ti_{94-x}Nb_xZr_6$ alloys plotted as a function of Nb-content and temperature, $E_H = f(T, x_{\mathrm{Nb}})$. The bottom left corner corresponds region of instability. Dashed and solid lines indicate to the borders of dynamical and mechanical instability, respectively.

In Figure 8, we compare calculated $E_H$-modulus at $T$=300 K with some available experimental data on Ti-Nb-Zr-based alloys [78-83]. Our simulations underestimate the $E_H$ by 5 GPa compared to experimental data for Ti-xNb-4Zr and Ti-xNb-8Zr Ref. [78], however there is a good agreement between theoretical and experimental trends of concentration dependence of elastic modulus. Some of Ti-Nb-Zr-alloys, which we consider as an experimental reference, contain small amount of other elements, so we cannot make a direct comparison with our theoretical results. However, we see that MLIP calculations of $E_H$-modulus can be reliable, and overall, there is a good agreement between our results and available experimental data on Ti-Nb-Zr-based alloys.

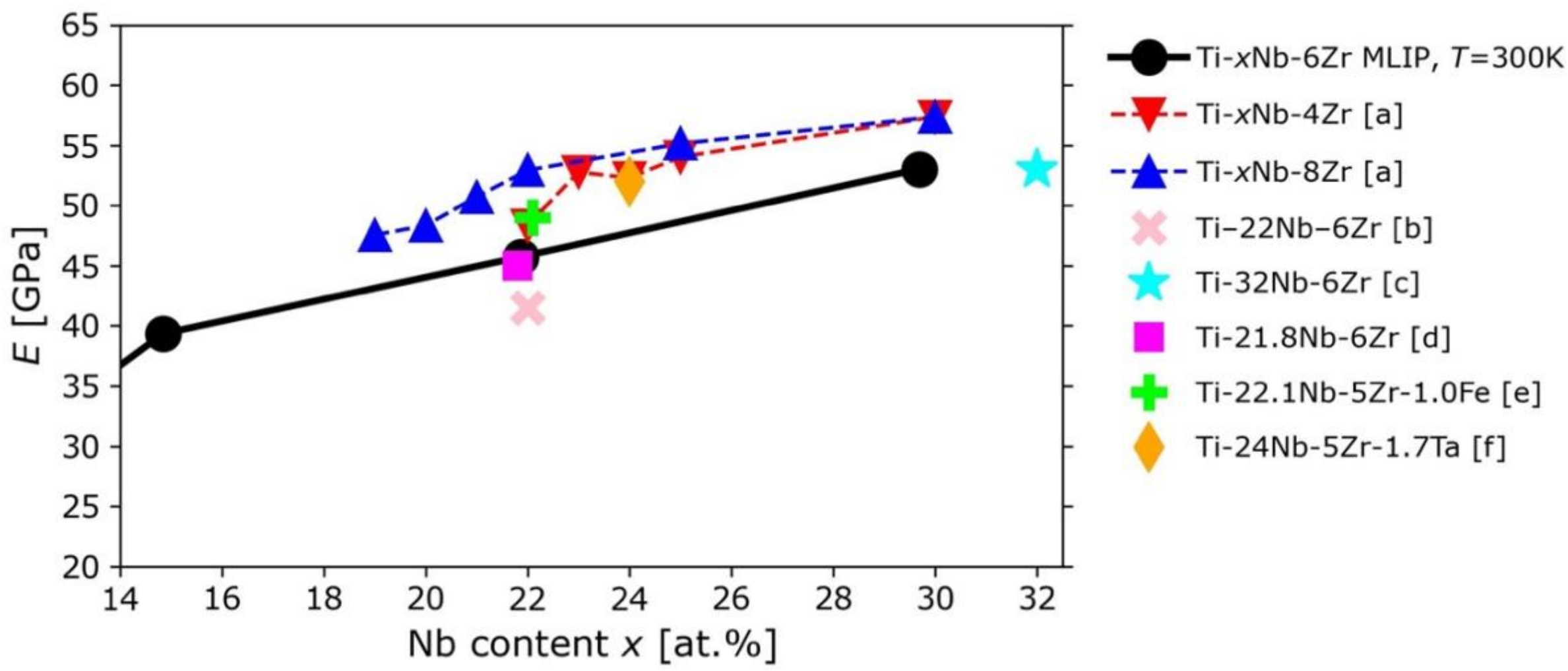

Figure 8. Comparison of our simulations of Young's modulus $E_H$ of Ti-Nb-Zr alloys at $T$ = 300 K calculated in this work (solid line with filled circles) with available experimental data on Ti-Nb-Zr-based alloys (at.%): Refs. [a-f] – [78-83].

### 3.5 Anisotropy of elastic modulus

Previous studies carried out at T = 0 K have shown that Ti-V alloys in vicinity of mechanical instability exhibit significant differences in moduli when determined using the Reuss and Voigt methods [54]. In our work, similar trend is observed for $Ti_{94-x}Nb_xZr_6$ alloys even at elevated temperatures. Figure 9 shows the Young's modulus of these alloys as determined by the Voigt, Reuss and Hill method at 300 K and 500 K. At 300 K and low Nb content, there is a notable difference between $E_R$ and $E_V$, indicating a pronounced anisotropy in the directional $E$-modulus (see Figure 9a). As Nb content increases, the difference between $E_R$ and $E_V$ becomes smaller, as well as the anisotropy of the directional $E$-modulus.

As previously noted, higher temperatures improve the mechanical stability of alloys. At 500 K, the difference between $E_R$ and $E_V$ is less significant than at 300 K, particularly in the $Ti_{88}Nb_6Zr_6$ alloy. With increasing temperature, the anisotropy also decreases, as shown in the directional $E$-modulus results in Figure 9b. For all the studied alloys, the highest directional $E$-modulus was found along the [111] crystallographic direction, while the lowest was along the [100] direction.

Pilz et al. [84] reported that experimental β-Ti-42Nb (at.%) alloys demonstrate strong anisotropy in directional Young's modulus, with the highest and lowest measured values of 79 ± 3 GPa and 44 ± 2 GPa, respectively. The maximum directional Young's modulus was observed along the [111] crystallographic direction, and the minimum along the [100] direction [84]. Tane et al. [85] also observed similar anisotropy in experimental Ti-29Nb-Ta-Zr and Ti-25Nb-Ta-Zr samples: the highest and lowest Young's modulus were measured along [111] and [100] direction, respectively. While we cannot directly compare our calculations with these experimental data, there is a good qualitative agreement between these results.

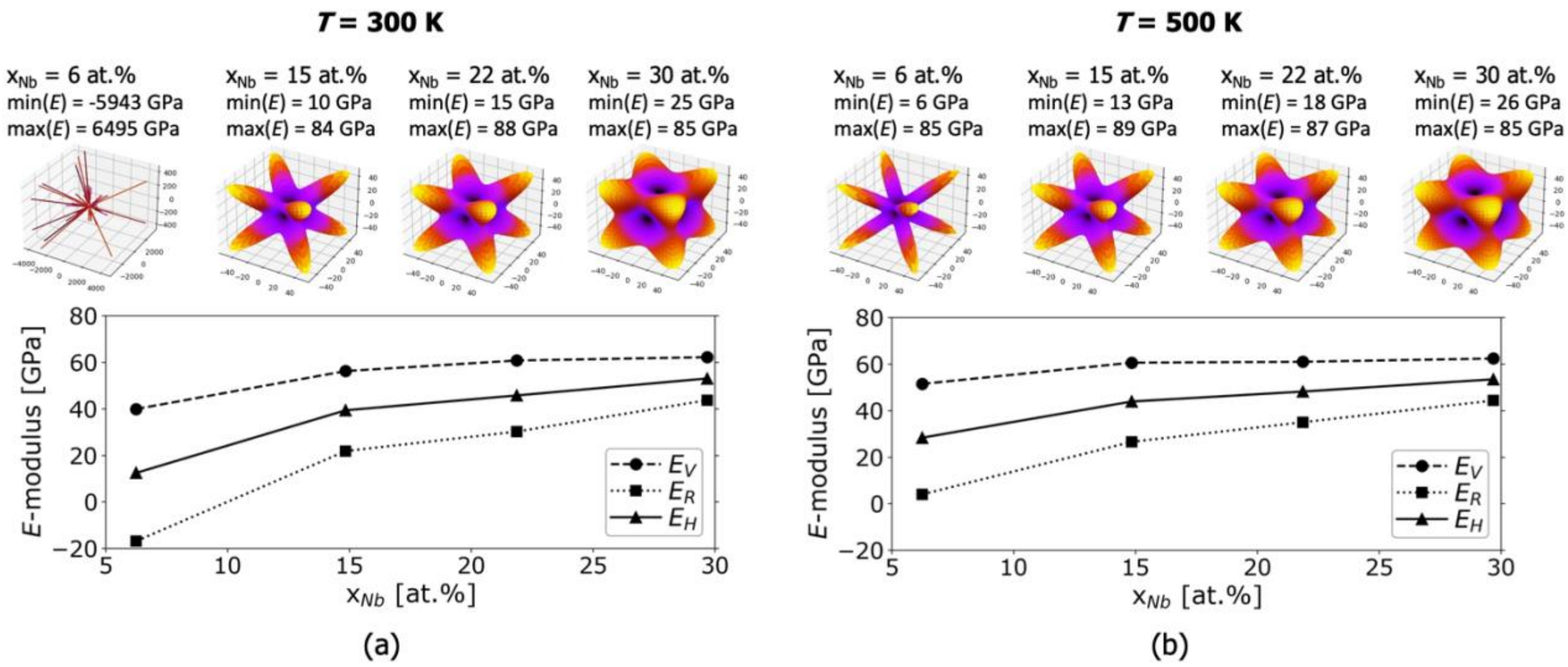

Figure 9. Young's modulus of $Ti_{94-x}Nb_xZr_6$ alloys calculated using Voight, Reuss and Hill methods at temperatures (a) 300 and (b) 500 K. 3D-directional Young's moduli with minimum and maximum values are shown for each composition.

## Conclusion

We addressed the challenging task of simulating the material properties near the point of dynamical instability. In this work, we demonstrated that machine learning interatomic potential can serve as a powerful tool for this purpose. Trained potential accurately predicts the elastic properties of β-type $Ti_{94-x}Nb_xZr_6$ alloys. These alloys exhibit notable signs of dynamical and mechanical instability at low Nb concentrations and low temperatures.

Near the point of dynamical/mechanical instability, we see a softening of Young's modulus of alloys. Furthermore, the anisotropy of directional Young's modulus becomes stronger while approaching the dynamical/mechanical instability. This finding highlights a possibility to design alloys with desired mechanical properties by tailoring the texture of materials in a vicinity of dynamical and/or mechanical instability. Specifically, by combining the information on anisotropy of directional Young's modulus with softening of Ti-Nb-Zr alloys in the vicinity of instability, it is possible to optimize alloy compositions to achieve a low elastic modulus comparable to that of human bone.

## Acknowledgements

Computations were enabled by resources provided by the National Academic Infrastructure for Supercomputing in Sweden (NAISS) at NSC partially funded by the Swedish Research Council through Grant Agreement No. 2022-06725. We acknowledge support from the Knut and Alice Wallenberg Foundation (Wallenberg Scholar grant no. KAW-2018.0194) and the Swedish Government Strategic Research Area in Materials Science on Functional Materials at Linköping University (Faculty Grant SFOMat-LiU No. 2009 00971).

**Data availability**

The data supporting the finding of this study is available in Open Materials Database under reference number [86]. It includes the SQS structurers for Ti-Nb-Zr alloys and the dataset for training the MTP potential.

**Supplemental information for:**

# Machine learning interatomic potential for the low-modulus Ti-Nb-Zr alloys in the vicinity of dynamical instability

Boburjon Mukhamedov[1], Ferenc Tasnádi[1] and Igor A. Abrikosov[1]

[1]*Theoretical Physics Division, Department of Physics, Chemistry and Biology (IFM), Linköping University, SE-581 83, Linköping, Sweden*

## S1. Effect of cutoff distance on the performance of MLIP

We evaluated the performance the MTP trained with different values of cutoff distance: 5 Å, 6 Å, and 7 Å. For this we used our dataset that contained 880 configurations after the active learning. First, we shuffled the dataset and split it into a training set (800 configurations) and a testing set (80 configurations). We then trained three separate MTPs with different cutoff radii

Figures S1 and S2 show the performance of these MTPs on the training and testing sets. The scatter plots compare MTP-predicted total energies and stress components against the corresponding DFT values. Based on these comparisons, we conclude that the MTP with a cutoff radius of 5 Å performs well, and increasing the cutoff does not significantly improve the accuracy of the potential. Additionally, we observed that training the MTP with a 7 Å cutoff required approximately 2.5 times more CPU core-hours compared to the 5 Å cutoff, further reinforcing the trade-off between computational cost and accuracy.

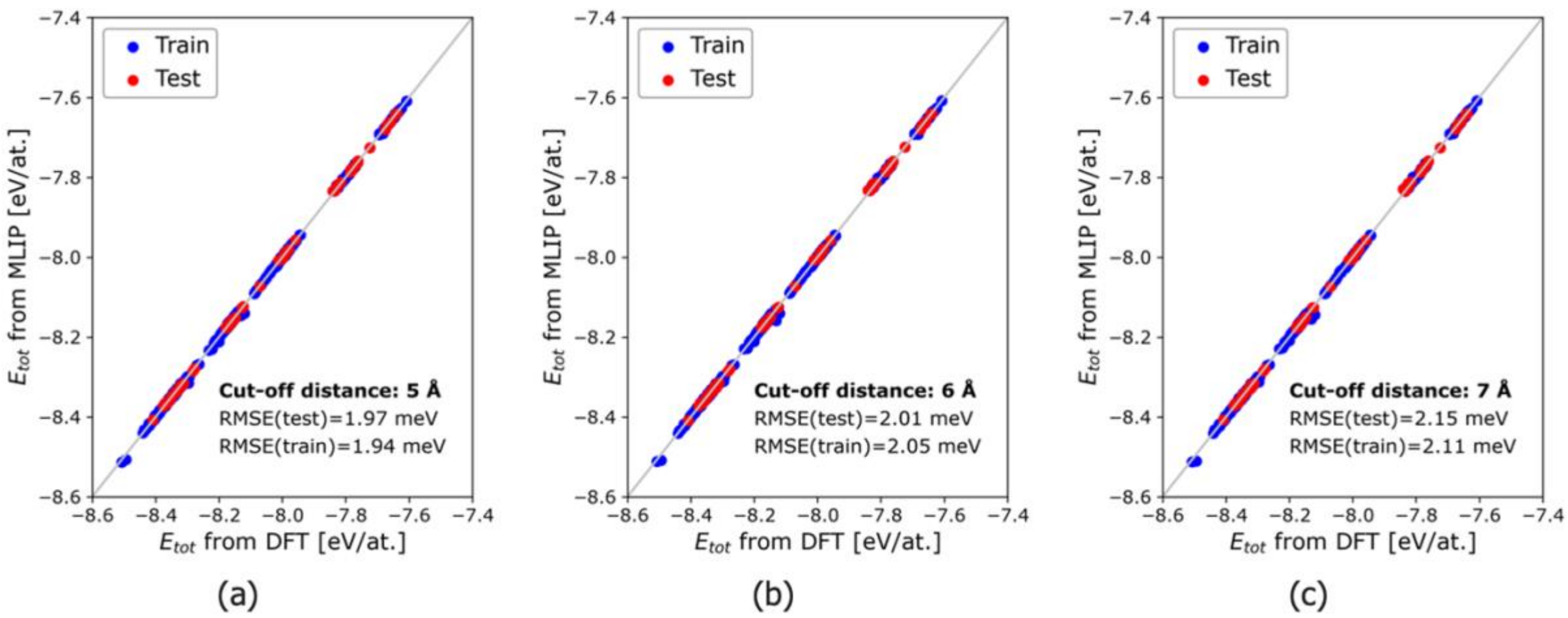

Figure S1. Scatter plots of MTP-predicted total energies versus DFT-calculated total energies for Ti-Nb-Zr alloys. The three panels correspond to MTPs with different cutoff distances: 5 Å, 6 Å, and 7 Å. The plots include both training and testing datasets, with the corresponding RMSE values of the MTP predictions.

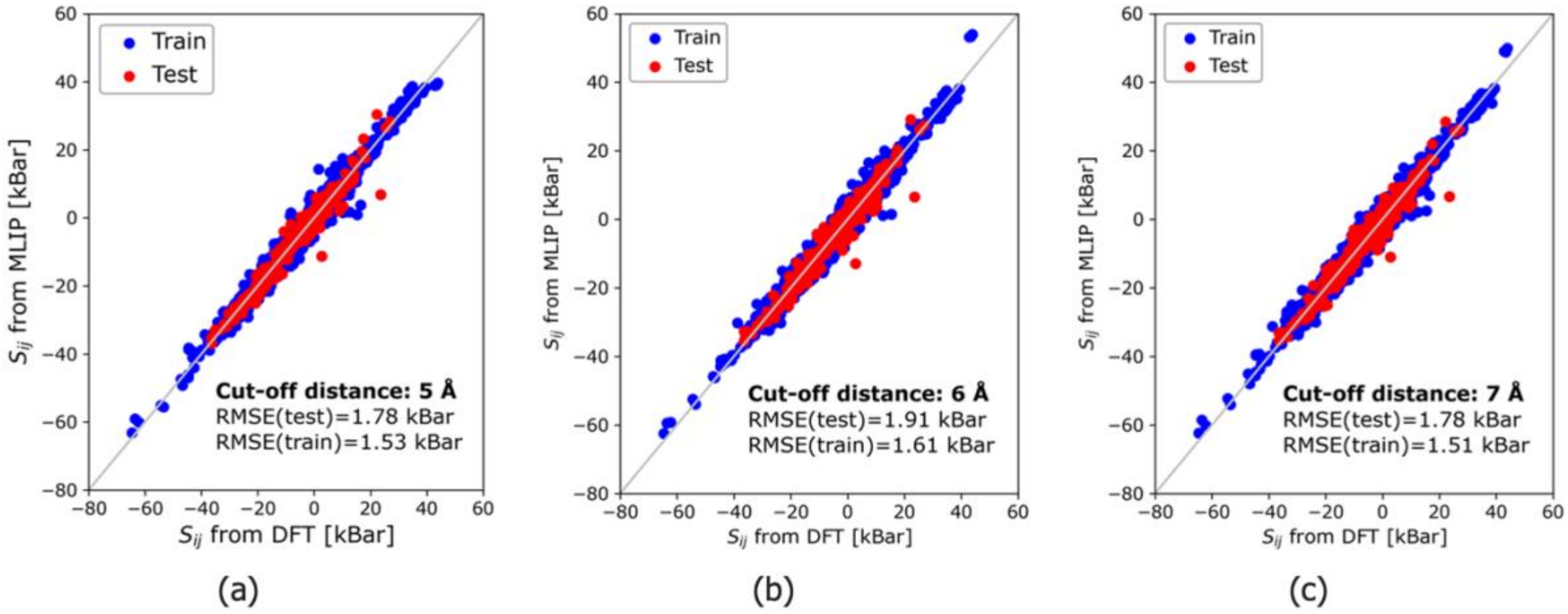

Figure S2. Scatter plots of MTP-predicted stress components versus DFT-calculated stresses for Ti-Nb-Zr alloys. The three panels correspond to MTPs with different cutoff distances: 5 Å, 6 Å, and 7 Å. The plots include both training and testing datasets, with the corresponding RMSE values of the MTP predictions.

## S2. Performance of MLIP in the regions of dynamical instability

To assess the performance of our MTP in regions of dynamical instability in Ti-Nb-Zr alloys, we generated an independent test set that was not used during training. We began by conducting **NPT-MD simulations** for **$Ti_{93}Nb_1Zr_6$** and **$Ti_{88}Nb_6Zr_6$** at **300 K and 500 K**. Each simulation ran for **15,000 time steps**, with MD configurations saved in four separate "dump" files.

From each dump file, we randomly selected **10 configurations**, excluding the first **5,000 time steps**, and performed **static DFT calculations** on these structures. In total, the test set consisted of **40 new configurations**, labeled as the **"Test" set**.

**Figure S3** illustrates the performance of the MLIP on this new test set. It compares **MTP-predicted total energies and stress components** with corresponding **DFT values**, and also presents the RMSE for the test set. These results demonstrate that the trained MTP provides **accurate predictions** of atomic-scale properties relevant to MD simulations, even at **low temperatures and low Nb concentrations**.

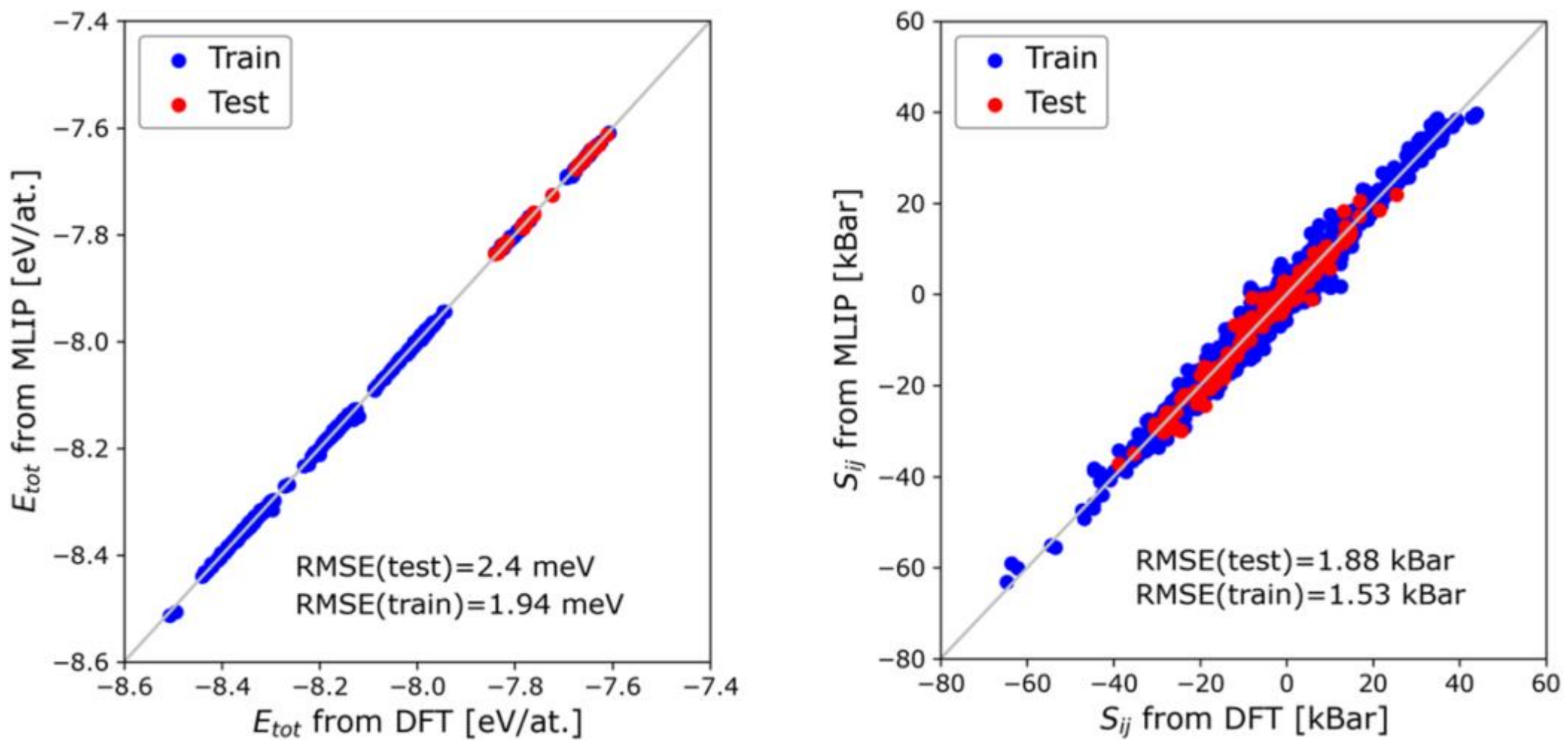

Figure S3. Scatter plots of MTP-predicted (a) total energies and (b) stresses versus DFT-calculated values for Ti-Nb-Zr alloys. "Test" set denotes the independent data for low Nb alloys which was not included in the training set.

**S3. Average structures of $Ti_{94-x}Nb_xZr_6$ SQSs**

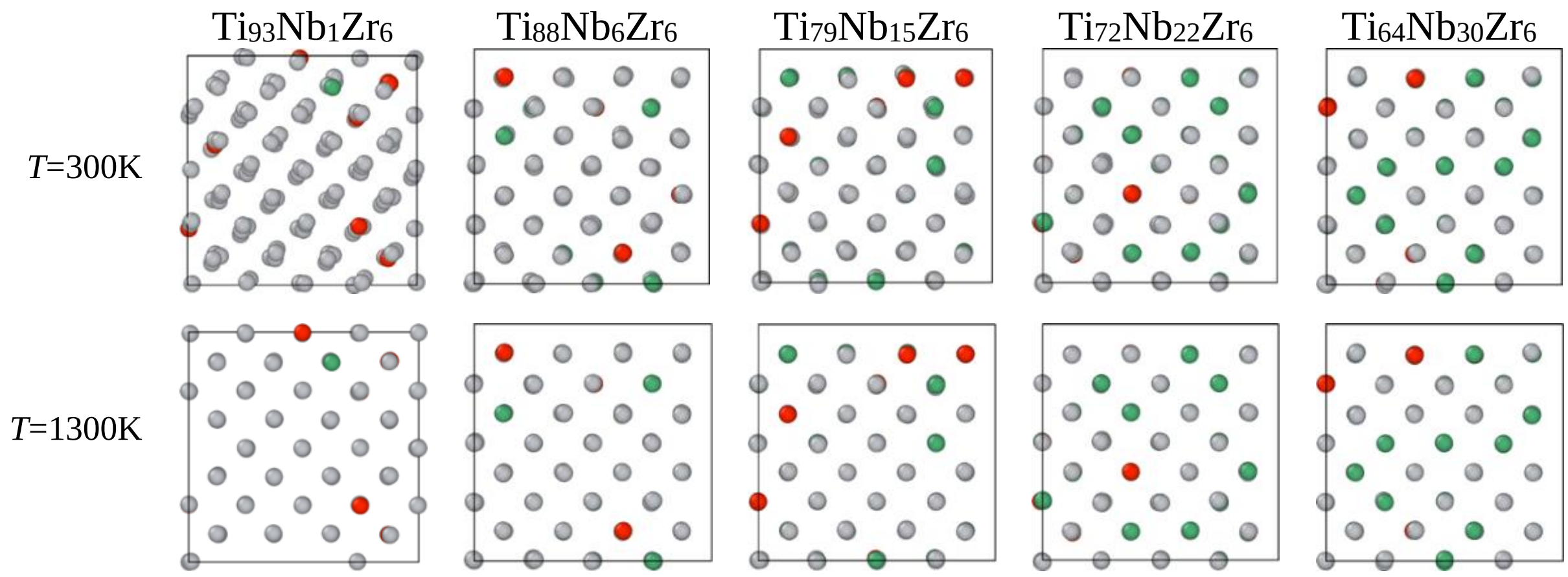

Figure S4. Average structures of five $Ti_{94-x}Nb_xZr_6$ SQSs derived from MD simulations conducted at 300 K and 1300 K. The averaging process included configurations from timestep 5,001 to 30,000, with the initial 5,000 timesteps omitted from analysis. The structures shown in the upper row represent the results at 300 K, while those in the lower row correspond to 1300 K.

## S4. Atomic displacements in $Ti_{94-x}Nb_xZr_6$ SQSs after ionic relaxation at *T* = 0 K.

Figure S5 presents the (100) projections of DFT-calculated structures after relaxation of interatomic forces at 0 K and zero pressure. For low-Nb compositions, significant atomic displacements from ideal bcc positions were observed, suggesting that the relaxed structures could not be classified as bcc. In contrast, high-Nb alloys ($Ti_{79}Nb_{15}Zr_6$, $Ti_{72}Nb_{22}Zr_6$, and $Ti_{64}Nb_{30}Zr_6$) mostly maintained bcc symmetry after relaxation, albeit with slightly increased atomic displacements in $Ti_{79}Nb_{15}Zr_6$ (see Figure S5 (c) and (f)).

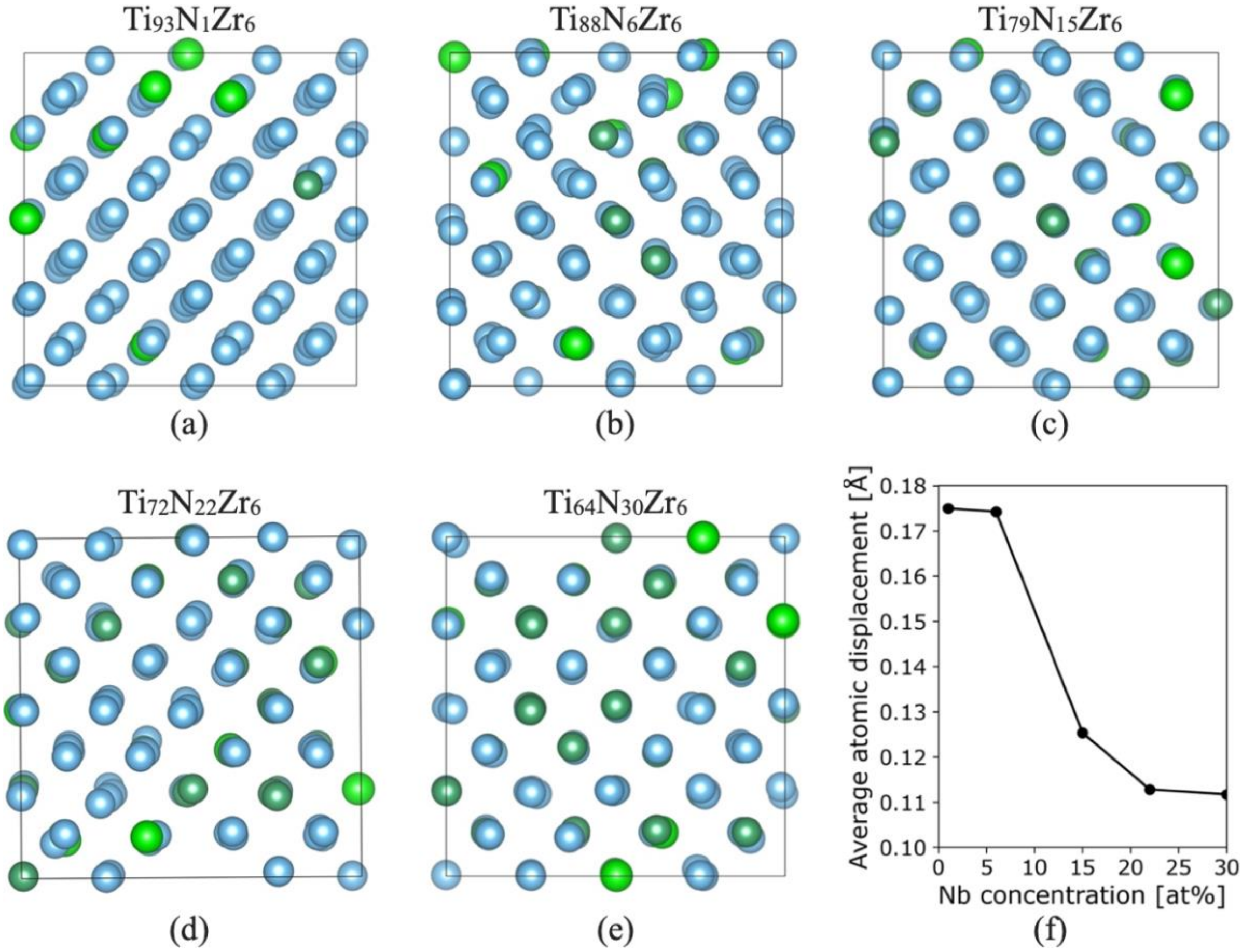

Figure S5. The (100) projections of DFT-calculated structures for (a–e) $Ti_{94-x}Nb_xZr_6$ SQSs after relaxation of interatomic forces at $T = 0$ K and zero pressure; (f) Average atomic displacement from ideal bcc positions as a function of Nb concentration in $Ti_{94-x}N_xZr_6$.

## S5. Comparison of MLIP and DFT results for unrelaxed SQS configurations at 0 K.

Figure S6 and Table S1 demonstrate comparison between DFT and MLIP predictions for $C_{ij}$ elastic constants, lattice parameters, and total energies in unrelaxed SQS configurations. In this calculation, we minimized the total energy and determined the equilibrium volume without performing ionic relaxation, ensuring that the bcc symmetry remained intact. MLIP overestimates $C_{12}$ and underestimates $C_{44}$. The discrepancies between DFT and MLIP values are approximately: $\Delta C_{11} \approx 2$ GPa, $\Delta C_{12} \approx 5-6$ GPa and $\Delta C_{44} \approx 3-4$ GPa.

MLIP slightly underestimates lattice parameters at 0 K, though the difference from DFT predictions is less than 0.1%. The total energy differences between DFT and MLIP predictions at 0 K are approximately 2-4 meV/atom. Overall, we see a good agreement between DFT and MLIP in predicting $C_{ij}$ constants and other properties at 0 K.

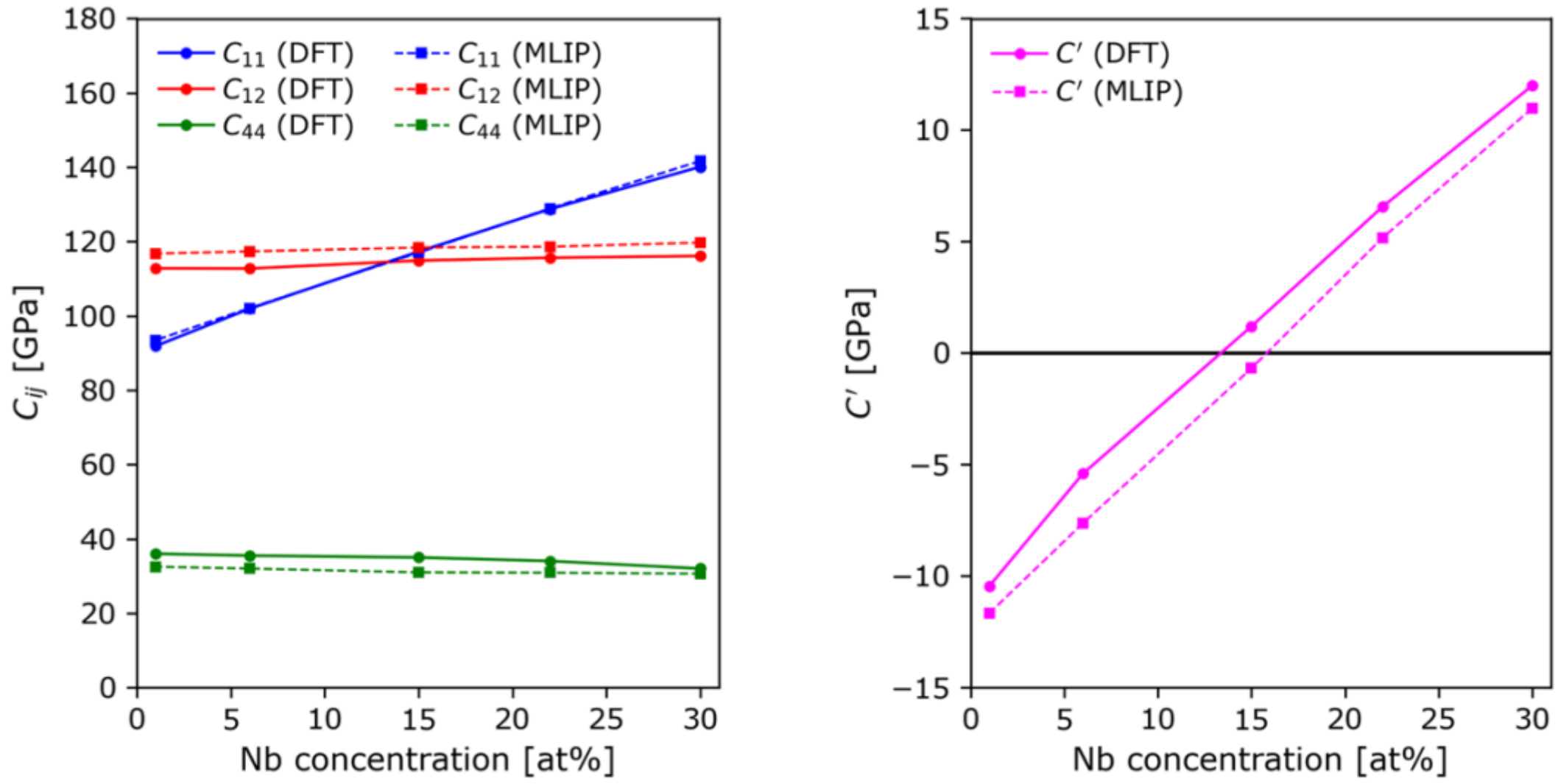

(a) (b)

Figure S6. (a) Elastic constants $C_{ij}$ for unrelaxed SQSs and (b) $C'$-parameters of alloys at $T$ = 0 K. Solid and dashed lines correspond to DFT-calculated and MLIP-predicted results, respectively.

Table S1. Lattice parameters and total energies of unrelaxed $Ti_{94-x}N_xZr_6$ alloys at $T$ = 0 K. Only the volumes of SQSs were relaxed to meet a zero pressure requirement.

| ***x*, at.%** | **Lattice parameter $a$, Å** | | | **Total energy $E_{tot}$, eV/atom** | | |
|---|---|---|---|---|---|---|
| | DFT | MLIP | $\Delta a$ | DFT | MLIP | $\Delta E_{tot}$ |
| 1 | 3.2749 | 3.2718 | 0.0030 | -7.7738 | -7.7762 | 0.0024 |
| 6 | 3.2763 | 3.2732 | 0.0031 | -7.9125 | -7.9153 | 0.0028 |
| 15 | 3.2786 | 3.2764 | 0.0022 | -8.1280 | -8.1312 | 0.0032 |
| 22 | 3.2813 | 3.2793 | 0.0020 | -8.3042 | -8.3077 | 0.0034 |
| 30 | 3.2853 | 3.2830 | 0.0023 | -8.4982 | -8.5019 | 0.0037 |

## S6. MSD in smaller supercells

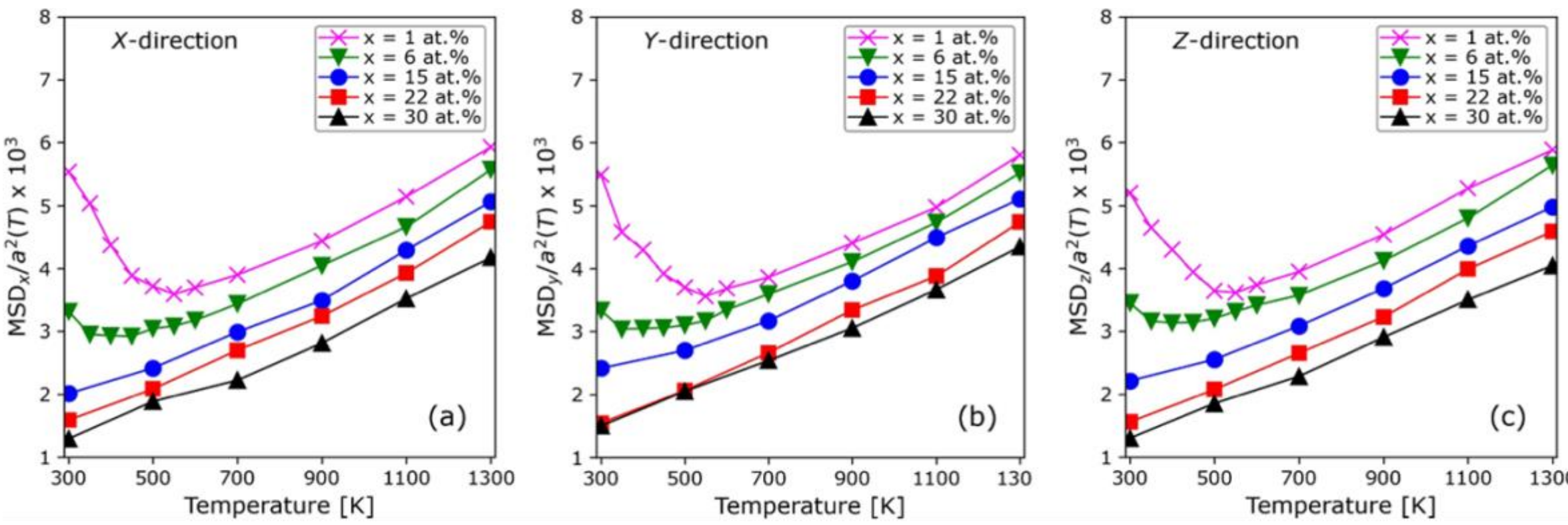

Figure S7. Temperature dependence of mean square displacements MSD of atoms from their ideal lattice positions in $Ti_{94-x}Nb_xZr_6$ alloys. The MSDs were calculated for supercells containing 128 atoms and were determined along three crystallographic directions: (a) $MSD_X$, (b) $MSD_Y$ and (c) $MSD_Z$. The MSDs were determined within NVT simulations, accounting for the thermal expansion of the alloys.

**S7. Strain matrix. Full elastic tensor. Averaging of $C_{ij}$**

Elastic constants $C_{ij}$ of Ti-Nb-Zr alloy SQSs were calculated using the stress-strain relation. The strain values were set to ±2 and ±4 %. Compared to perfect cubic structures the SQSs exhibit lower symmetry, therefore one needs to calculate the full elastic tensor as shown:

$$C_{ij} = \begin{bmatrix} C_{11} & C_{12} & C_{13} & 0 & 0 & 0 \\ C_{21} & C_{22} & C_{23} & 0 & 0 & 0 \\ C_{31} & C_{32} & C_{33} & 0 & 0 & 0 \\ 0 & 0 & 0 & C_{44} & 0 & 0 \\ 0 & 0 & 0 & 0 & C_{55} & 0 \\ 0 & 0 & 0 & 0 & 0 & C_{66} \end{bmatrix}$$

Then, the average values of $C_{11}$, $C_{12}$ and $C_{44}$ constants were determined as:

$$\bar{C}_{11} = \frac{C_{11} + C_{22} + C_{33}}{3}$$

$$\bar{C}_{12} = \frac{C_{12} + C_{13} + C_{23}}{3}$$

$$\bar{C}_{44} = \frac{C_{44} + C_{55} + C_{66}}{3}$$

$$C' = \frac{\bar{C}_{11} - \bar{C}_{12}}{2}$$

Further, we use only those average values of $C_{ij}$ for analysis of the elastic properties of Ti-Nb-Zr alloy.

**S8. Voight-Reuss-Hill averaging for polycrystalline moduli**

Below we provide the equations for the polycrystalline moduli determined using Voight (V), Reuss (R) and Hill (H) averaging methods.

Bulk modulus:

$$B_R = B_V = B_H = \frac{C_{11} + 2C_{12}}{3}$$

Shear modulus:

$$G_R = \frac{5 \cdot C_{44} \cdot C'}{2 \cdot C_{44} + 3 \cdot C'} \qquad G_V = \frac{3 \cdot C_{44} + 2 \cdot C'}{5} \qquad G_H = \frac{G_R + G_V}{2}$$

Young's modulus:

$$E_R = \frac{9 \cdot B_R \cdot G_R}{3 \cdot B_R + G_R} \qquad E_V = \frac{9 \cdot B_V \cdot G_V}{3 \cdot B_V + G_V} \qquad E_H = \frac{E_R + E_V}{2}$$